\documentstyle[12pt,aasms4,psfig]{article}

\begin{document}

\title{The Bull's-Eye Effect as a Probe of $\Omega$}

\author{Adrian L. Melott\altaffilmark{1}
	Peter Coles\altaffilmark{1,}\altaffilmark{2},
        Hume A. Feldman\altaffilmark{1},
	and
        Brian Wilhite\altaffilmark{1}}

\altaffiltext{1}{Department of Physics and Astronomy, University of
        Kansas, Lawrence, KS  66045, USA}
\altaffiltext{2}{Astronomy Unit, Queen Mary \& Westfield College,
University of London, London E1 4NS, UK}

\begin{abstract}
We compare the statistical properties of structures normal and
transverse to the line of sight which appear in
theoretical N-body simulations of structure formation, and
seem also to be present in observational data from
redshift surveys.  We present a statistic which can quantify this
effect in a conceptually different way from standard analyses of
distortions of the power-spectrum or correlation
function. From tests with $N$--body experiments, we argue
that this statistic represents a new and potentially powerful diagnostic
of the cosmological density parameter, $\Omega$.
\end{abstract}


\section{Introduction}

According to the Hubble law,
$v\simeq cz \simeq H_0 d$, the recession velocity $v$ of a galaxy,
inferred from its redshift, is
proportional to its proper distance from the observer, $d$. Irregularities
gravitationally generate peculiar velocities
($v_p$), so that the true relationship is $v=H_0 d + v_p$,
where $v_p$ is the line of sight component of the
peculiar motion. Maps of galaxy
positions constructed by assuming that velocities are exactly
proportional to distance (redshift space) have two principal distortions. The
first is generated in dense collapsed structures where there
are very many galaxies at essentially the same distance from the observer,
each with a random peculiar motion.
This results in a radial stretching
of the structure known as a ``Finger
of God''. The second effect
acts on much larger scales (e.g. Kaiser 1987).
A large overdensity generates coherent bulk motions in the galaxy distribution
as
it collapses. Material generally flows towards the center of the structure,
i.e. towards the observer for material on the far side of the structure and
away from the observer on the near side so it will appear compressed along
the line of sight.
This tends to make voids bigger and the walls between them denser.

One particularly important unresolved issue
is  the value of the cosmological density parameter,
$\Omega$, which determines whether the universe will expand forever
($\Omega < 1$),  or eventually recollapse ($\Omega > 1$).
See e.g. Coles and Ellis (1997) for a review.
While redshift-space distortions
are a nuisance when one wants to construct
accurate maps of the (true) spatial distribution of galaxies,
they  may lead to a robust determination of $\Omega$.
Unfortunately, applications of this idea to existing data
have so far yielded inconclusive results for $\Omega$
(Hamilton 1998 and references therein)
and recent
studies indicate that systematic errors in the standard approaches
are likely to be less manageable than has previously been supposed
(e.g. Hatton \& Cole 1997; see also de Laix and Starkman 1997).

We propose here an entirely new approach to the analysis of these
distortions, based on the statistics of the spacing of features (Praton
\& Schneider 1994; Praton, Melott \& McKee 1997) both orthogonal and tangential
to the observer's line of sight in redshift space: the ``Bull's-eye
Effect.'' Praton et al. (1997) showed that preferentially concentric redshift
space features emerging in N--body simulations were dynamical in origin
with only very small additional contributions from survey geometry or
magnitude limits.  Simulations with no peculiar velocities or with
randomized velocities show no enhanced Bull's--eye pattern.
We show herein a method of quantifying this difference
in the redshift direction, which depends on large--scale motions.
This approach bypasses much of the bias dependence implicit in the
standard methods of analysis, possibly leading to a much cleaner
determination of the density parameter $\Omega$.
Further, this effect exists independent of the selection function.
A selection function may mildly enhance the visual prominence of the effect,
but it is created by redshift distortions.

Concentric rings seem to appear in
such surveys as Las Campanas (Schectman et. al 1996) and CNOC2
(Yee et al. 1997).
We have developed a
quantitative method to detect this signal,
which is useful in distinguishing between high and low $\Omega$.
If peculiar velocities are
low, and observed concentric structures are not enhanced in
redshift space (as advocated for the Great Wall by Dell'Antonio et. al
1996), this is not a refutation of the Bull's-Eye effect.  It is
evidence for low $\Omega$.

\section{Dynamics of The Bull's-Eye Effect}

The effect of peculiar motions is to increase the spacing of
large--scale structures in the radial direction as compared with real
space, as we show quantitatively below. The enhancement is
$\Omega$--dependent.  It is on the basis of this visually-striking
difference between the redshift-space behavior of low-density and
high-density dynamics that we propose a statistic that reproduces the
eye's sensitivity to differences in pattern.

It is easier to introduce and explain this effect in the small--angle
approximation to redshift space. A distant patch of the Universe will
have nearly parallel radii and perpendicular to it circular arc segments
of constant $z$.
In Figure 1, we show an array of plots from such a Cartesian N--body
simulation,
described fully in Beacom et. al (1991). Upper left and right represent
sequential stages. The two lower plots show the stage at the upper left viewed
along its $y$ or $x$ axes in redshift space. There is considerable large--scale
displacement in this $\Omega = 1$ model, which acts to somewhat anticipate
structures that will appear later in real space (upper right). Figure 2 shows
the same thing in an $\Omega_0 = 0.2$ simulation.  While the ``Fingers of God''
are still present, the large--scale displacements are largely gone.

The essence of large--scale redshift space effects is a compression
and/or expansion effect along the line of sight.  It can best be
explained using the Zel'dovich approximation (Zel'dovich 1970).
This is now known to
reproduce weakly non-linear (i.e. large--scale) features in the
distribution of matter very accurately indeed, if implemented in an
optimized form known as the Truncated Zel'dovich Approximation
(Coles et al. 1993, Melott 1994). This
approximation follows the development of structure by relating the final
(Eulerian) position of a particle ${\bf r}$ at some time $t$ to its
initial (Lagrangian) position ${\bf q}$ defined at the primordial epoch
when particles were smoothly distributed:
$$
{\bf r}= a(t) {\bf x}({\bf q},t)= a(t)[{\bf q} - D_+(t) \nabla_{\bf q}
\Phi ({\bf q})],
\eqno(1)
$$
where $D_+(t)$ describes the linear growth of perturbations
as a function of cosmological proper time, $t$, and
$a(t)$ is the cosmic scale factor (c.f. Peebles, 1980). In the case
of a flat universe, $D_+(t)= a(t)$; the behavior of
$D_+$ for more general cosmologies is usually parameterized by
$f=d\log D_+/d\log a$, where $f$ can be accurately approximated as
$\Omega^{0.6}$.

In the separable mapping (1), the displacement field is
given by the gradient of the primordial gravitational potential $\Phi$
with respect to the initial coordinates.
Differentiating the expression (1) leads to
$$
{\bf V}= {d {\bf r}\over dt} = H{\bf r}-a(t)\dot{D}_+ \nabla_{\bf q}
\Phi({\bf q})
\eqno(2)
$$
for the velocity of a fluid element ${\bf V}$.
The mapping (2) provides a straightforward explanation of the changed
characteristic scale of structures in the redshift direction.
Calculating the redshift coordinate exactly and translating it into an
effective distance $d_z$ gives
$$
d_z={V\over H}=r_3 - f a(t)D_+(t)
\nabla_3 \Phi ({\bf q})=aq_3-(1+f) a(t)D_+(t) \nabla_3 \Phi ({\bf q}),
\eqno(3)
$$
in which we have the 3-axis in the redshift direction.
(Note that $\delta$, the density contrast, does not enter here.) Thus the
displacement term becomes multiplied by a factor $(1+f)$ in (3) compared
to (2).

The effect of the displacement field in redshift space is to
give the observer a ``preview'' (albeit in only one direction) of a
later stage of the clustering hierarchy. In effect, the characteristic
scale of structures in the z-direction should be larger than that
in the directions unaffected by the extra displacements induced
by redshift-space distortions. We interpret the characteristic
scale as the typical distance between {\em caustics}
where the mapping (1) becomes singular, which are the prominent
high-density features of large-scale structure.

At first sight, this argument seems to suggest that the characteristic
the characteristic spacing of high-density regions should be
a factor of $(1+f)$ higher for the z-direction than in the orthogonal
direction, i.e. a factor of two difference in a critical density universe.
The real situation is, however, more complicated than that.  The
dominant effect of the extra displacements in (3) compared to
(1) is their tendency to compress structures together,
decreasing their number per unit length and increasing the typical space
between them.  This effect is, however, contaminated by
the Fingers of God, which tend to increase the apparent radial extent
of high-density regions and therefore push neighboring structures
together, even in low-density universes where the large-scale
displacement effect is negligible. The upshot of this is that,
for a critical density universe, the effect is rather less than
the naive factor of two but, as we shall see, there is still
a highly significant difference in characteristic spacings
between low-- and high--$\Omega$ models.

This effect is rather more subtle than the simple amplification of the
power spectrum amplitudes one obtains by applying Eulerian perturbation
theory to the problem (Kaiser 1987). The change in pattern from
real to redshift space is due to phase changes associated with
the extra displacements. These changes do not appear in the power
spectrum (which ignores phase information),
but are visually clear as they result in the merging of high-density
structures and a consequent increase in the spacing between them.

\section{Quantifying Displacements}

Our method
has the character of a first-order statistic (it is related to the mean
spacing of objects) rather than the usual second-order statistics (such
as the power spectrum), which are blind to the number-density of objects.
First, we smooth the distribution in the
simulations to produce a continuous density field in order to
erase strongly non-linear small-scale structures.
We have found empirically that smoothing
by a Gaussian $\exp (-r^2/\alpha^2 r_0^2)$ where $r_0$ is defined by
$\xi(r_0)=1$ and $\alpha=0.5$ gives the
best results, with weak sensitivity to smoothing scale.
We then construct density contours
for the smoothed field, and
take lines-of-sight through the smoothed density field and calculate the
{\it rms} distance between successive
up-crossings of the {\it same} contour level in the radial direction;
denoted $S_{\parallel}$. Note that there will be one down-crossing of the
same level between
any two successive up-crossings. Since it is our intention to capture caustics,
we concentrate on high-density contours.
In fact, we found that for a variety of $\Omega$ and power spectral values,
the most reliable approach is to choose a contour level by specifying
the filling factor of regions above the contour, not an absolute
$\delta$, and that the level corresponding to a filling factor of
15\%, about the peak of the spacing, is an excellent choice.

We also do a similar calculation for tangential lines (in the direction
orthogonal to the observer's line of sight); this is denoted $ S_{\perp}$.
After much experimentation
with a large ensemble of simulations, one statistic turned out to be
nearly optimal:
the ratio of the {\em rms} spacing in the redshift direction to that in the
orthogonal direction, which we call $\mu$:
$$
\mu ={ S_{\parallel}\over S_{\perp}}
\eqno(4)
$$
We verified that $\mu$ is independent of the power spectrum, and amount
of nonlinearity, for a given background cosmology. We tested a total of
16 different high resolution 2d simulations -- four realizations
each, power law spectra $P(k)\propto k^2$, or constant, in both low (0.2) and
high ($\Omega=1$) models.
The {\it rms} spacing for an $\Omega=1$ universe is
larger in the redshift direction than in the direction orthogonal to the
line of sight. As $\Omega_0\rightarrow 0$, the ratio decreases.

In Figure 3a, we show the {\it rms} distances $S$ as a function of filling
factor. Boldface denotes those measured in the redshift direction, light the
orthogonal one; low $\Omega_0$ is dashed, high $\Omega$ has solid lines.  It is
clear that redshift space intervals are larger than real space ones.
(We defer showing error bars to 3b, to avoid a crowded plot.)
Is the ratio $\Omega$--dependent? We show $\mu$ in 3b, with one-$\sigma$
errors in the mean obtained by averaging over all lines of sight
in an ensemble of N-body simulations.
These errors are typically larger than the measurement and systematic
errors of the redshift.
It is clear that $\mu$ is much larger in a critical density
Universe--at about a 4$\sigma$ confidence level.

We have also performed a preliminary analysis of a pair of large (256$^3$)
simulations with a CDM-type power spectrum in a low-- and critical--density
background. In these simulations we have used the
spherical symmetry that would be seen by a real observer, rather than
the simplified Cartesian redshift space shown in the Figures for
illustration. However, a full study of this type will require a variety of
analyses with varying initial power spectra, amplitudes, survey geometries and
magnitude limits which is beyond the scope of this Letter.

\section{Discussion}

We have described a new statistic for
extracting estimates of $\Omega$ from redshift space distortions.
We demonstrated that this method is very sensitive to the choice of
$\Omega$.  We have described its dynamical origin, and showed that it
is qualitatively different than other related statistical methods.
More detailed and systematic
tests of the method are necessary,
against an ensemble of simulations in
a bigger variety of background cosmologies with realistic geometry and
selection functions.
It is appropriate here to mention that we do not expect our method
to be sensitive to the presence of a cosmological constant.
However, an extension of the method may
just make it possible to measure $q_0$.
Our method has picked out the radial anisotropy due to redshift distortion,
which could not be picked out with high significance by examining the
orientation of voids in redshift space diagrams (Ryden and Melott 1996).
In extremely deep redshift surveys, variation of $\mu$ with redshift may
provide an independent estimate of $q_0$.

Note that $\mu$ is the ratio of a displacement to a displacement. Unlike
the usual methods, we do not relate displacement (or velocity) to
$\delta$, which is the origin of bias--dependence (e.g. Hamilton 1998,
\S 4.1). This is confirmed by the relatively weak dependence of $\mu$
on the filling factor chosen to select contours in our method.
Dynamically-produced voids resemble flat-bottomed valleys surrounded
by steep mountains. A local bias has the effect of moving the contour level
up the mountains, but will not lead to a large increase in the typical
size of the valleys unless the contour level changes so much that
the filling factor is reduced by a large factor.
Since a large bias does not seem to be allowed by standard
theories of structure formation, we do not expect bias to affect
$\mu$ significantly in realistic situations.
These considerations lead us to conclude that our technique is expected
to be significantly less  sensitive to the presence of a bias $b$ than the
standard methods. The main effect would be that we would choose
too large a value of the smoothing length (for $b>1$) for the
actual state of dynamical evolution of the distribution.
We have already found experimentally that this has a much
weaker effect than $b$ has in the usual second-order methods, such as the
power spectrum and correlation function.

Of course, it remains  possible that some extreme form of bias might
occur which does affect the reliability of our method. For example, it
is clear that a galaxy distribution that was simply ``painted'' onto
a uniform high-density background (e.g. Bower et al. 1993) with
no dynamical tracer of its origin, would fool our method (as it would the
standard tools). If the bias were linear but so large that it completely
swamped the effects of dynamical evolution we would also expect to have
problems: the mean spacing of high-density
regions in such models could  be much larger than any scale associated
with dynamics and $\mu$ might well turn out to be very close to unity
even if $\Omega_0=1$. We are not claiming, therefore, that our method
is completely independent of all possible forms of biasing, but we
are confident that a modest amount of linear bias does not have the
same adverse effect on our technique as it does on standard methods.
We are currently using larger three-dimensional numerical simulations
to explore this effect in detail.

A different and probably more significant limitation of our approach is that
it will require a  catalog with a relatively high
number-density of galaxies so that excessive smoothing is not
required. The sampling rate of the Las Campanas Survey (Shectman et
al. 1996) may be adequate for this purpose, but the method is ideally
suited to upcoming very large surveys, particularly the Anglo-Australian
2DF and Sloan Digital Sky Survey.

\acknowledgements

We wish to acknowledge the support of the NSF-EPSCoR program,
NASA grant NAG5-4039,
as well as the Visiting Professorship, General, and Undergraduate
Research Funds of the University of Kansas. Peter Coles receives a PPARC
Advanced Research Fellowship.
We are also grateful to
Sergei Shandarin, Beth Praton, and Barbara Ryden for useful conversations. An
anonymous referee was very helpful in improving the presentation.

\clearpage
%
\begin{figure}
\begin{center}
  \leavevmode\psfig{figure=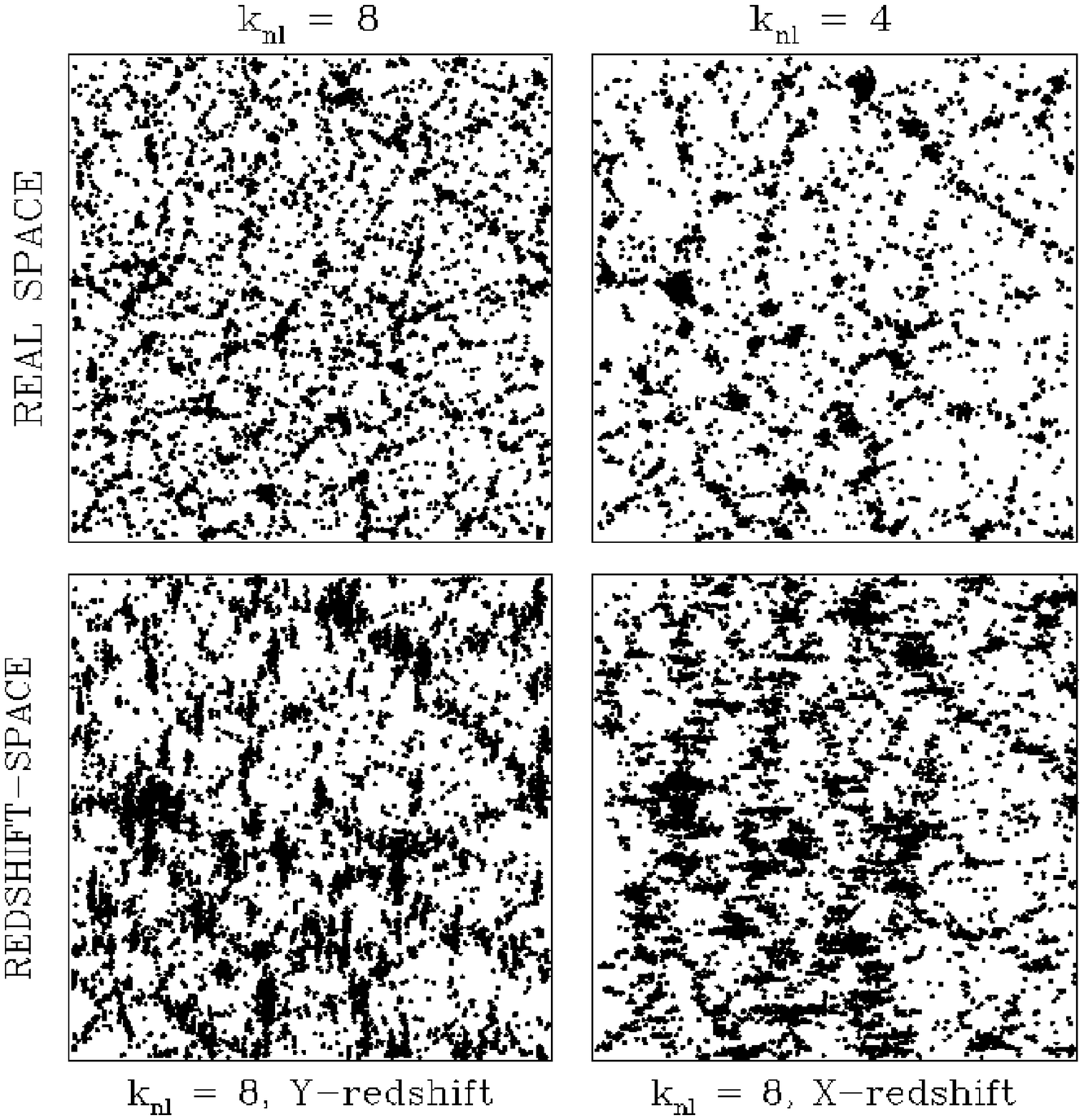,height=16cm}
\end{center}
\caption{
Top-left: a map showing the true positions of all particle positions
in a two-dimensional simulation with $\Omega=1$ and an initial
power spectrum $P(k)\propto k^{0}$ (corresponding to $n=-1$ in 3D).
The top right panel shows the same realization evolved to an expansion
factor $a(t)$ twice that of the preceding panel. The bottom two
panels show the top-left panel viewed in redshift space
using the $y$ and $x$ directions (small angle distant observer approximation).
Note the increased spacing between
structures along the redshift direction.
\label{fig:fig01}}
\end{figure}
%
%
\begin{figure}
\begin{center}
  \leavevmode\psfig{figure=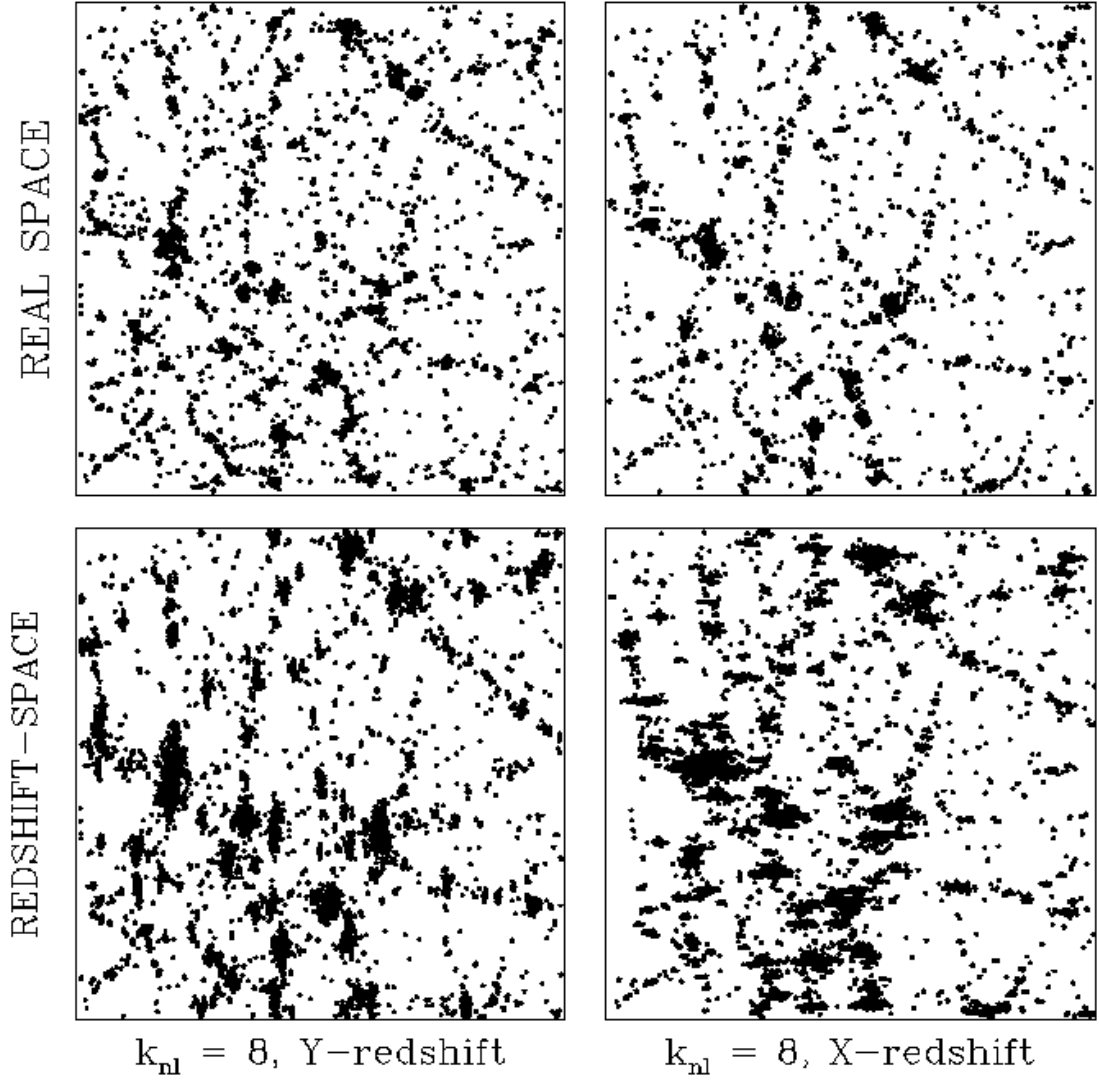,height=16cm}
\end{center}
\caption{
The same as in Figure 1, but for a simulation with $\Omega_0$ = 0.2.
\label{fig:fig02}}
\end{figure}
\begin{figure}
\begin{center}
  \leavevmode\psfig{figure=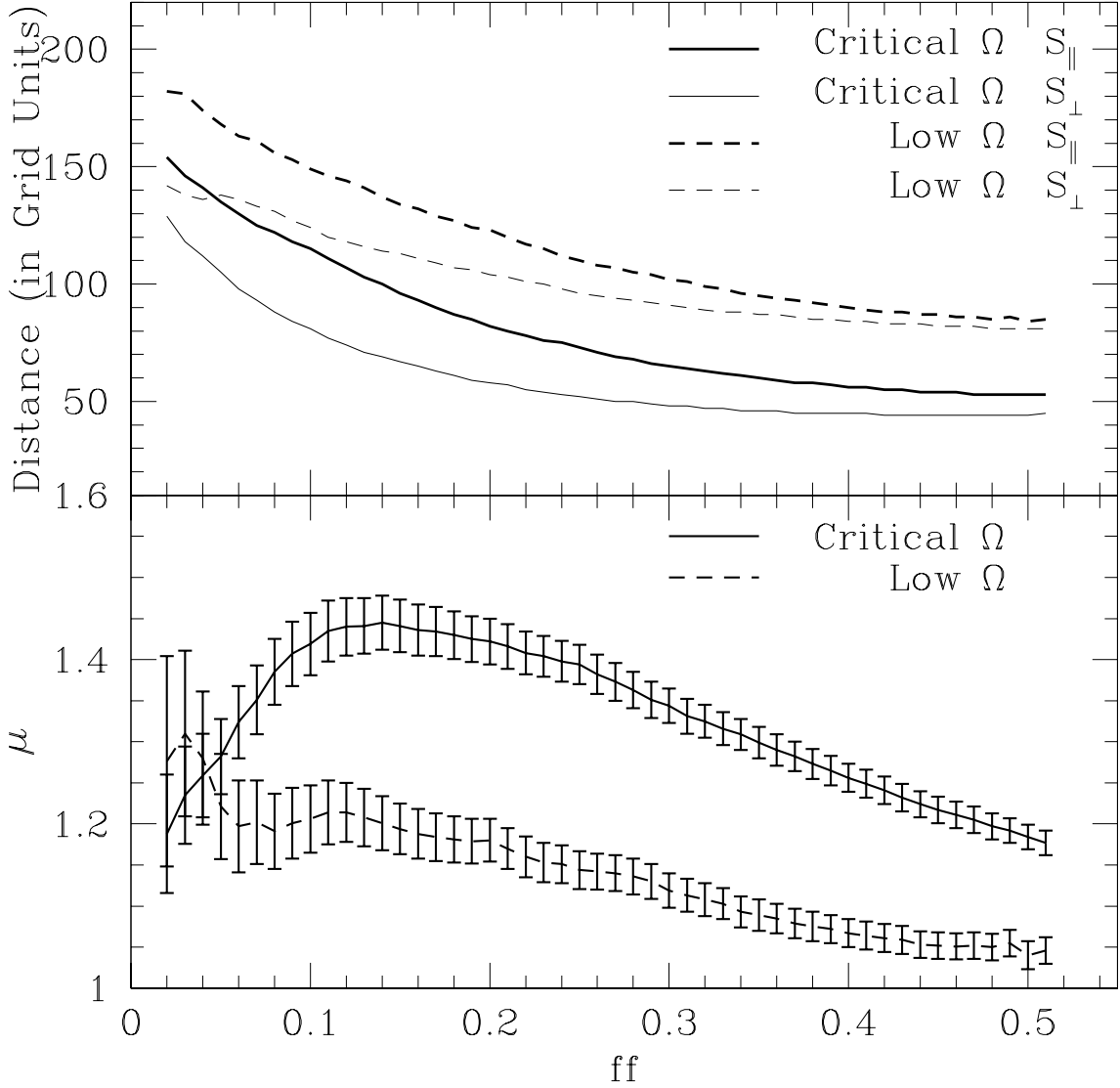,height=16cm}
\end{center}
\caption{(a) The {\it rms} spacing of contour crossings
(in computer grid units in which the simulation box has length
512) in the redshift (bold lines)
and tangential (light) directions for the $\Omega=1$ simulations (solid lines)
and low--$\Omega$ simulations (dashed lines) for different contour levels
as a function of filling factor (the fraction of space above the threshold).
(b) The measure $\mu$, which is the ratio of the {\it rms}
redshift space distance and the {\it rms} distance in the orthogonal direction,
confirms the visual impression of Figures 1 and 2.
\label{fig:fig03}}
\end{figure}

\end{document}